\documentclass[sigconf]{acmart}

\usepackage{graphicx}
\usepackage{textcomp}
\usepackage{tcolorbox}
\usepackage{pifont}
\usepackage{xcolor}
\usepackage{soul}
\usepackage{booktabs}
\usepackage{multirow}
\usepackage{float}
\usepackage{url}
\usepackage{balance}
\usepackage{microtype}
\usepackage[numbered]{bookmark}
\usepackage{titlesec}
\usepackage{seqsplit}
\usepackage{newunicodechar,graphicx}
\usepackage[utf8]{inputenc}
\usepackage{pgf-pie}  
\usepackage{balance} 
\usepackage{algorithmic}
\usepackage{array}
\usepackage{tabularx} 
\usepackage{enumitem}
\setlist[itemize,enumerate]{noitemsep, topsep=0pt, leftmargin=1.0em}
\setlist{nolistsep}
\usepackage[utf8]{inputenc}
\usepackage{newunicodechar,graphicx}
\DeclareRobustCommand{\okina}{%
  \raisebox{\dimexpr\fontcharht\font`A-\height}{%
    \scalebox{0.8}{`}%
  }%
}
\newunicodechar{ʻ}{\okina}
\newcommand{\rev}[1]{\textcolor{black}{#1}}

\newcommand{\RQA}{\textbf{RQ1}: How do developers perceive the importance of accessibility in mobile apps?}

\newcommand{\RQB}{\textbf{RQ2}: What accessibility features do developers implement in their mobile apps, and what challenges impede their integration?}

\newcommand{\RQC}{\textbf{RQ3}: How do developers approach accessibility testing in mobile apps?}

\newcommand{\RQD}{\textbf{RQ4}: What factors prevent developers from implementing accessibility features in their mobile apps, and what changes would enable adoption?}

\newcommand{\RQE}{\textbf{RQ5}: How do accessibility practices vary by platform specialization and developer experience?}

\copyrightyear{2026}
\acmYear{2026}
\setcopyright{cc}
\setcctype{by}
\acmConference[ICSE '26]{2026 IEEE/ACM 48th International Conference on Software Engineering}{April 12--18, 2026}{Rio de Janeiro, Brazil}
\acmBooktitle{2026 IEEE/ACM 48th International Conference on Software Engineering (ICSE '26), April 12--18, 2026, Rio de Janeiro, Brazil}
\acmPrice{}
\acmDOI{10.1145/3744916.3787791}
\acmISBN{979-8-4007-2025-3/2026/04}

\begin{document}

\title[Practitioner Views on Mobile App Accessibility: Practices and Challenges]{Practitioner Views on Mobile App Accessibility:\\ Practices and Challenges}

%%
%% The "author" command and its associated commands are used to define
%% the authors and their affiliations.
%% Of note is the shared affiliation of the first two authors, and the
%% "authornote" and "authornotemark" commands
%% used to denote shared contribution to the research.
\author{Amila Indika}
\email{amilaind@hawaii.edu}
\orcid{0000-0003-3379-7047}
\affiliation{%
  \institution{University of Hawaiʻi at Mānoa}
  \state{Hawaiʻi}
  \country{USA}
}

\author{Rick Kazman}
\email{kazman@hawaii.edu}
\orcid{0000-0003-0392-2783}
\affiliation{%
  \institution{University of Hawaiʻi at Mānoa}
  \state{Hawaiʻi}
  \country{USA}
}

\author{Anthony Peruma}
\email{peruma@hawaii.edu}
\orcid{0000-0003-2585-657X}
\affiliation{%
  \institution{University of Hawaiʻi at Mānoa}
  \state{Hawaiʻi}
  \country{USA}
}

\begin{abstract}
As mobile applications (apps) become ubiquitous in everyday life, it is crucial for developers to prioritize accessibility for users with diverse abilities. While previous research has identified widespread accessibility issues and raised awareness of developer challenges, there remains a lack of cross-platform, globally representative insights into how practitioners approach accessibility in practice. This paper presents findings from a mixed-methods survey of 110 mobile app developers across 43 countries, examining \rev{how platform ecosystems (iOS vs. Android) and developer experience shape accessibility practices.}
Results show that while developers recognize the importance of accessibility, they often rely on platform-specific guidelines and typically perform compliance testing late in the development process. Developers primarily implement text-focused features while struggling with API limitations and organizational constraints. 
\rev{Through systematic cross-platform comparison, we identify novel platform-specific barriers and demonstrate how accessibility practices differ across developer experience levels.
Our findings offer new insights into the challenges of achieving accessibility in practice and provide actionable steps for various stakeholders to promote more consistent and inclusive app development.}
\end{abstract}

\flushbottom
\raggedbottom
\maketitle

\section{Introduction}
Mobile applications (apps) have become an integral part of daily life, providing users with easy access to information and services. With over 2 million apps available across multiple platforms \cite{web-appStoreCount}, users have a wide selection of apps to choose from for various purposes, including communication, games, health, and finance \cite{web-appStoreCategory}. This growth and diversity of mobile apps is driven by technological advancements that enable individuals and small businesses to create apps without needing formal training in software engineering or related fields \cite{Xanthopoulos2013, Heitktter2013, Nunkesser2018, Azmy2023, walter2015learning, Oltrogge2018}.

While this democratization of app development is a positive trend, it raises concerns about the potential consequences of inexperienced developers entering the field. Limited awareness of best practices results in low-quality apps with maintenance issues, poor performance, security vulnerabilities, and negative user experiences. These shortcomings can frustrate users, incur costs, and harm the reputation of apps and developers \cite{AbuFarha2024}.

Among these best practices is incorporating accessibility features to support people with disabilities. According to the World Health Organization, 16\% of the global population (i.e., 1.3 billion) %(i.e., 16\% of the global population) 
experience significant disabilities \cite{WHO_disability_report}, highlighting the need for inclusive design that addresses barriers preventing disabled users from accessing services and information \cite{hersh2019learning}. To help developers create accessible software, organizations and governing bodies have established standards and guidelines, such as the Web Content Accessibility Guidelines (WCAG) \cite{guidelines-wcag}, ISO 9241-171 \cite{guidelines-ISO9241171}, and ETSI EN 301 549 \cite{guidelines-EN01549}, as well as mobile-specific guidelines from Apple and Google \cite{guidelines_ios, guidelines_android, guidelines_ios2, guidelines_android2}.

\rev{
Although much research has examined mobile app development, most studies have focused on Android, as evidenced by numerous systematic reviews \cite{Li2017, Senanayake2023, Zhan2022, Fawad2024}. This single-platform focus overlooks important insights due to the unique development paradigms of Android and iOS apps, which differ in technical implementation, their developer ecosystems, methodologies, user experiences, and release strategies \cite{Joorabchi2013, Heitktter2013, Ahmad2017, Ali2017, Hu2018, Domnguezlvarez2019}. For example, iOS development relies on Apple's proprietary tools and frameworks, such as Xcode and Swift, within a stringent ecosystem. In contrast, Android development leverages open-source technologies, such as Java and Kotlin, which enable compatibility across diverse devices and operating system versions. These differences require distinct approaches to development, testing, and deployment for each platform \cite{Joorabchi2013}.
}

\rev{Furthermore, current research on mobile app accessibility remains limited. \cite{Paiva2021,Bi2022}.} Prior research, based on repository mining, has revealed widespread accessibility issues in apps, such as missing labels, inadequate text contrast, and undersized elements \cite{chen2021accessible, yan2019current, ross2020epidemiology}. While these technical analyses provide valuable insights, they primarily focus on Android and overlook developers' perspectives. Similarly, survey-based approaches have shown that developers lack awareness of accessibility principles \cite{alshayban2020accessibility, di2022making}. However, these surveys typically feature limited sample sizes and focus on specific regions or platforms. \rev{Importantly, no prior research has empirically investigated how developer experience relates to accessibility practices. This includes examining how differences in ecosystems influence the implementation of accessibility features in apps, as well as when and how developers test for accessibility.}

To understand the practices and challenges of incorporating accessibility features into mobile apps, it is essential to engage directly with the primary stakeholders (i.e., app developers) 
\cite{Lenberg2015}. We thus adopt a socio-technical approach, examining both the technical aspects and human factors of software development, which provides richer insights than purely technical analyses \cite{Storey2020, Stol2018}. By engaging with a diverse, global group of app developers, our study presents \rev{a cross-platform empirical perspective on the} real-world practices and challenges of implementing accessibility\rev{, revealing how experience level and ecosystem constraints influence accessibility design, implementation, and testing}.

\subsection{Goal and Research Questions}
\label{subsection: research_questions}
Our study is a comprehensive survey of \rev{cross-platform} mobile app developers across various regions to understand the practical aspects of mobile app accessibility from a developer's perspective. We investigate how developers perceive the importance of accessibility, the features they implement, their testing approach, and the barriers they encounter \rev{when building apps for multiple platforms}. \rev{Understanding platform-specific practices and barriers is vital, as differences between ecosystems, such as technical constraints, marketplace policies, and device fragmentation, influence how accessibility is prioritized and implemented in practice. But a lack of accessibility awareness across platforms can lead to inconsistent end-user experiences, which disproportionately impact users with disabilities. Our} insights can help vendors, organizations, and educators create more effective platform-specific accessibility tools, policies, and training programs.
Hence, we address the following research questions (RQs):

\vspace{1mm}
\noindent\textbf{\RQA} This RQ examines developers' perceptions of app accessibility. By understanding their priorities regarding accessibility, we can better understand why accessibility features may be overlooked or deprioritized in development workflows.

\vspace{1mm}
\noindent\textbf{\RQB} This RQ focuses on the specific accessibility features that developers incorporate into mobile apps and the barriers that make such features difficult to implement. The findings provide insight for vendors, researchers, and organizations to develop solutions that overcome these barriers.

\vspace{1mm}
\noindent\textbf{\RQC} This RQ aims to understand how developers ensure that their app meets accessibility standards, helping to identify gaps in current testing approaches, and the means to support developers in improving their apps' accessibility.

\vspace{1mm}
\noindent\textbf{\RQD} This RQ identifies the barriers preventing accessibility implementation and determines what changes would motivate greater adoption. By understanding these obstacles and potential solutions, strategies can be developed to promote and support the integration of accessibility practices into app development.

\vspace{1mm}
\noindent\textbf{\RQE} This RQ focuses on how accessibility practices evolve with developer experience and mobile platform. This understanding enables the creation of targeted resources, training, and strategies to support developers in advancing their accessibility practices.

\subsection{Contributions}
This study advances the field of mobile accessibility through: 
\rev{
\begin{itemize}
    \item \textbf{A cross-platform, global scope study.} An empirically grounded comparison of accessibility practices among mobile app developers, specifically for iOS and Android, based on survey responses from 110 practitioners across 43 countries. 
    \item \textbf{Comprehensive insights into accessibility testing.} An analysis of when, how, and by whom accessibility testing is conducted throughout the software development lifecycle, highlighting common practices and missed early testing opportunities. 
    \item \textbf{Developer experience analysis.} An examination of how developer experience levels and their familiarity with accessibility guidelines influence their implementation and testing behaviors.  
    \item \textbf{A systematic categorization of barriers.} An identification of socio-technical challenges spanning technical, organizational, and ecosystem levels that hinder accessibility adoption, providing a means for designing targeted interventions by tool vendors, educators, and platform providers.
\end{itemize}
}

\section{Study Design}
\label{Section:method}

\begin{table*}[!htbp]
\centering
\caption{Below are the questions that are part of the survey. The questionnaire, as presented to participants, which includes the answer options for the single-choice questions (SCQ) and multi-choice questions (MCQ), is available at \cite{ArtifactPackage}.}
\vspace{-3mm}
\label{Table:SurveyQuestions}
\resizebox{\linewidth}{!}{%
\begin{tabular}{
>{\centering\hspace{0pt}}m{0.03\linewidth}|   % No.
>{\hspace{0pt}}m{0.66\linewidth}|            % Survey Question
>{\raggedright\hspace{0pt}}m{0.09\linewidth}|  % Type
>{\hspace{0pt}}m{0.24\linewidth}             % Notes
}

\hline
\multicolumn{1}{>{\centering\hspace{0pt}}m{0.032\linewidth}|}{\textbf{No.}} & 
\multicolumn{1}{>{\centering\hspace{0pt}}m{0.606\linewidth}|}{\textbf{Survey Question }(* indicates answer required)} & 
\multicolumn{1}{>{\centering\hspace{0pt}}m{0.098\linewidth}|}{\textbf{Type}} & 
\multicolumn{1}{>{\centering\arraybackslash\hspace{0pt}}m{0.192\linewidth}}{\textbf{Notes}}  \\ 
\hline

1 &
  Do you consent to participate in this study? * &
  Yes/No &
  End Survey if ``No" \\ \hline
2 &
  Do you have experience in building and/or maintaining mobile apps? * &
  Yes/No &
  End Survey if ``No" \\ \hline
3 &
  How do you describe yourself? * &
  SCQ &
  \\ \hline %Includes ``Prefer to self-describe" free-text option
4 &
  In which country do you currently reside? * &
  SCQ &
   \\ \hline
5 &
  How many years of general programming experience do you have? * &
  SCQ &
   \\ \hline
6 &
  What best describes your current employment status? * &
  SCQ &
  Includes free-text option \\ \hline
7 &
  How many years of mobile app development experience do you have? * &
  SCQ &
   \\ \hline
8 &
  To what extent is building mobile apps part of your job/employment duties? * &
  SCQ &
   \\ \hline
9 &
  Which mobile app distribution platforms do you utilize to share/distribute your apps? * &
  MCQ &
  Includes free-text option \\ \hline
10 &
  How did you initially learn mobile app development? * &
  SCQ &
  Includes free-text option \\ \hline
11 &
  In your opinion, how important is it for developers to add accessibility features to their mobile apps? * &
  SCQ &
   \\ \hline
12 &
  Please select which accessibility guidelines you are familiar with, if any. * &
  MCQ &
  Includes free-text option \\ \hline
13 & % original-> 14
  Which types of software have you implemented accessibility features for? * &
  MCQ &
  Includes free-text option \\ \hline
14 & % original-> 15
  Please select the accessibility features you have incorporated into mobile apps. * &
  MCQ &
  Shown if ``Mobile Applications" is selected for Q13 \\ \hline
15 & % original-> 16
  Other than the above accessibility features, what other features (if any) have you incorporated into your apps? &
  Free-Text &
  Shown if ``Mobile Applications" is selected for Q13\\ \hline
16 & % original-> 17
  Have you encountered challenges incorporating accessibility features into mobile apps? * &
  Yes/No &
  Shown if ``Mobile Applications" is selected for Q13 \\ \hline
17 & % original-> 18
  What are some key challenges you faced when implementing accessibility features into your mobile apps? * &
  Free-Text &
  Shown if ``Yes" is selected for Q16 \\ \hline
18 & % original-> 13
  If you have received feedback about accessibility issues with your apps, please let us know what common issues have been reported. &
  Free-Text &
  Shown if ``Mobile Applications" is selected for Q13 \\ \hline
19 &
  At what stages of the software development lifecycle do you test your mobile apps for accessibility compliance? * &
  Free-Text &
  Shown if ``Mobile Applications" is selected for Q13\\ \hline
20 &
  What techniques do you use to test the accessibility compliance of your apps? * &
  MCQ &
  Shown if ``Mobile Applications" is selected for Q13; Includes free-text option \\ \hline
21 &
  What, if any, changes or improvements would you like to see in accessibility-related tools and/or libraries/frameworks? &
  Free-Text &
  Shown if ``Mobile Applications" is selected for Q13 \\ \hline
22 &
  What resources do you use to learn about mobile accessibility standards and best practices or to get help with accessibility-related questions? * &
  Free-Text & 
  Shown if ``Mobile Applications" is selected for Q13 \\ \hline
23 &
  What are your reasons for not implementing accessibility features in mobile apps? * &
  Free-Text &
  Shown if ``Mobile Applications" is \textit{not} selected for Q13 \\ \hline
24 &
  What changes would need to occur for you to implement accessibility features in future mobile app development? * &
  Free-Text &
  Shown if ``Mobile Applications" is \textit{not} selected for Q13 \\

\hline
\end{tabular}
}
\end{table*}

Our research strategy employs a respondent-driven design using a sample survey methodology \cite{Stol2018}, in which data are collected from a diverse group of app developers via an online questionnaire. To ensure rigor, we adhered to the Consolidated Criteria for Reporting Qualitative Research (COREQ) guidelines \cite{Tong2007}, a framework for quality and transparency in qualitative research, which are recommended for qualitative software engineering research \cite{Lenberg2023}.

\subsection{Survey Design}
We used Qualtrics \cite{qualtrics} to design and publish the survey, ensuring only one response per participant. The survey consisted of 24 questions, covering participant consent, demographics, mobile app development experience, and familiarity with accessibility features. We designed the survey to address our research questions, outlined in Section \ref{subsection: research_questions}, and followed survey best practices \cite{kitchenham2002principles, kasunic2005designing, linaaker2015guidelines}. Table \ref{Table:SurveyQuestions} shows the questions that were part of our survey.

\subsection{Pilot Run}
As per best practices \cite{Kitchenham2002,linaaker2015guidelines}, we conducted a pilot run of the survey by contacting four mobile app developers from the authors' professional networks. Participants were asked to provide constructive feedback on the questionnaire, focusing on aspects such as clarity, logical flow, and readability. After reviewing their feedback, we made refinements to the survey that included correcting ambiguous phrasing, removing redundant questions, and reordering others.  

\subsection{Ethical Compliance}
Our study was approved by the Institutional Review Board \cite{Vinson2008}. Informed consent was obtained from all participants, and confidentiality was ensured by not collecting identifiable information. No compensation was provided to avoid coercion.

\subsection{Sampling Strategy and Recruitment}
Participants were recruited from LinkedIn \cite{linkedin_stats}, which has proven effective for recruiting skilled professionals for research studies \cite{mirabeau2013utility, kanij2013lessons, Peruma2024}. 
At the time of this study, LinkedIn's policies do not restrict researchers from contacting individuals for online surveys as long as they adhere to their rules and community standards \cite{linkedinPolicy}.%, provided this was done in accordance with their rules and community standards \cite{linkedinPolicy}.

We employed a purposive sampling approach to recruit participants with experience in mobile app development. Specifically, we utilized LinkedIn's search functionality to identify individuals whose profiles contained any of the following \rev{six search} terms: `mobile engineer,' `android engineer,' `ios engineer,' `mobile developer,' `android developer,' `ios developer.' These search terms were consistent with those used in a prior study that investigated mobile app security practices and challenges \cite{Peruma2024}. At the time of conducting this study, LinkedIn limited the number of search results to 1,000 \rev{(10 results per page) per search term}. 

For each search result, we selected 150 random individuals and manually reviewed their profiles to verify their current job title, work history, and mobile app development experience. Our focus was on identifying developers with hands-on app development experience, while excluding those with only academic knowledge. After finding potential participants, we sent connection requests\footnote{\rev{We initiated 900 connection requests, of which only 720 were accepted. Survey invitations were sent only to these 720 successful connections.}} and invited them to take our online survey via LinkedIn messaging. \rev{Since the survey was anonymous, we were unable to identify participants who did not complete it, so we could not send reminders.}

\subsection{Data Analysis}
We adopted a mixed-methods approach for data analysis, employing quantitative and qualitative methods \cite{wagner2020challenges}. The quantitative analysis involved statistical techniques to identify patterns and trends. For the qualitative analysis, we conducted a thematic analysis of open-ended responses. Our thematic analysis followed a structured process: familiarization, coding, theme identification, refinement, definition, and final analysis \cite{Braun2006}. Two researchers, who have extensive experience in mobile app development and research, independently coded the responses, documenting emerging patterns and resolving differences through discussion. \rev{Rather than calculating a quantitative measure of inter-rater reliability, we ensured robustness in our analysis through discussion until consensus was reached on the final themes. This negotiated agreement approach is recognized as an effective method for ensuring coding quality in qualitative research \cite{Armstrong1997, Campbell2013}.} We note that the responses often contained multiple themes. Additionally, some responses were unclear or irrelevant to the specific survey question and were excluded from the thematic analysis. Hence, the sum of responses across all identified themes for a question may not equal the total number of responses received for that question. \rev{Finally, while we noted recurring feature types in responses, the data was not enough to claim theoretical saturation. We aimed to identify categories reflecting the various accessibility features and practices participants report, rather than to create a detailed taxonomy.}

\section{Results}
\label{Section:results}
This section answers our RQs.\footnote{Responses from the pilot run were only used to identify issues with the questionnaire and are not included in our RQ results.}\textsuperscript{,}\footnote{Due to space limitations, we only report frequent observations in certain parts of this write-up. Our dataset includes the complete findings and is available at: \cite{ArtifactPackage}}  
We begin with the number of responses and participant demographics, followed by our RQ results.

\subsection*{\textbf{Survey Responses}}
Our survey, open from January 2024 to February 2025, received 163 responses. However, to ensure consistency in our analysis, we focused on survey responses from participants who answered all required questions. This approach enhances the reliability and validity of our findings, minimizing concerns related to incomplete data. After excluding incomplete responses, \textbf{110 responses} remained, which we used to address our RQs.

\subsection*{\textbf{Participant Demographics}}
To understand participant backgrounds, we gathered non-identifying demographic data through survey questions \#3 to \#10. Further, as per best practice, \cite{kasunic2005designing}, we positioned these fact-based demographic questions at the start of the survey, as they are typically easier to answer before more in-depth technical and subjective questions. We report the participants' demographic data in two sections: (1) general demographics and (2) mobile app development experience.

\noindent\textbf{General Demographics.}
The survey questions from \#3 to \#6 captured general demographics. The majority of participants identified as male (95 or 86.36\%), followed by female (14 or 12.73\%), with one individual self-describing (0.91\%). Regarding participants' countries of residence, responses included 43 nations. The distribution by continent included: Asia (39 or 35.45\%), Africa (28 or 25.45\%), Europe (19 or 17.27\%), North America (18 or 16.36\%), South America (4 or 3.64\%), and Oceania (2 or 1.80\%), representing a diverse sample.% This country distribution indicates a diverse, globally representative sample. 

Regarding general programming experience, the largest groups fell within the 6–10 years (46 or 41.82\%) and 3–5 years (44 or 40\%) ranges. Additionally, 13 participants (11.82\%) reported having over 10 years of experience, while only seven individuals had fewer than two years, suggesting that the participant pool mainly consists of experienced mobile developers. Further, most participants work full-time (96 or 87.27\%), while eight work part-time, and only six are full-time students, unemployed, or independent contractors.% (freelance), or immediate university graduates. 

\noindent\textbf{Mobile App Development Experience.}
Survey questions \#7 through \#10 examined participants' years of mobile app development experience, revealing that most had over three years of experience. Specifically, 47 participants (42.73\%) had 3–5 years of experience, 38 participants (34.55\%) had 6–10 years, and 10 participants (9.09\%) had over 10 years of experience. Only 15 participants had fewer than two years of experience, indicating that our study reflects insights from experienced mobile developers. Further, a majority of participants indicated that mobile development is a core part of their job responsibilities, with 66 participants (60\%) reporting they do it ``All of the time'' and another 29 (26.36\%) indicating they do it ``Most of the time''. Additionally, 10 participants (9.09\%) said ``A lot'', while two participants reported ``Somewhat'' and one reported ``Very little.'' Finally, two reported ``Not at all.''

The survey also examined the platforms participants used for app distribution via a multiple-choice question (MCQ). On average, participants use 1.65 platforms. The most common combination was Apple App Store and Google Play, chosen by 40 participants (36.36\%). Individually, Google Play was identified as the most frequently used platform (77 responses or 42.54\%), followed closely by the Apple App Store (75 responses or 41.44\%). Other distribution platforms mentioned included the Amazon App Store, Microsoft Store, Firebase Distribution, and Huawei AppGallery.

Regarding mobile development education, most participants (62 or 56.36\%) were self-taught through resources like books, podcasts, videos, and blogs. Only 17 (15.45\%) had formal education, 13 (11.82\%) attended coding boot camps or workshops, and 12 (10.91\%) completed online courses on platforms like edX or Pluralsight.

\vspace{1mm}
\noindent\textbf{\textit{Summary:}} 
The participant demographic shows a diverse sample across various regions and experience levels in programming and mobile development, predominantly comprised of industry professionals. Many participants developed for multiple platforms, strengthening the validity of our findings and offering insights into mobile accessibility practices in different global contexts.

\subsection*{\RQA}
This RQ examines developers' awareness and attitudes toward mobile app accessibility. We gathered these insights through survey questions \#11 and \#12, which assess the importance they place on accessibility and their familiarity with accessibility guidelines.

Survey question \#11 used a Likert scale to assess developers' perceived importance of mobile accessibility. %As shown in Figure \ref{fig: importance_of_accessibility}, 
Out of 110 participants, 54 (49.09\%) rated accessibility as ``Very important," 31 (28.18\%) as ``Moderately important," 17 (15.45\%) as ``Extremely important," and 8 (7.27\%) as ``Slightly important."

Survey question \#12, an MCQ, examined developers' familiarity with standard accessibility guidelines. A total of 45 participants selected multiple options, averaging 1.71 selections per participant. %As shown in Table \ref{tab: developers_accessibility_guideline_familiarity}, 
The most recognized guidelines were the Android Accessibility Guidelines, selected by 66 participants (35.11\%), followed by the Apple iOS Accessibility Guidelines, selected by 62 participants (32.98\%).  Further, 19 participants (10.11\%) reported familiarity with the Web Content Accessibility Guidelines (WCAG), while 15 participants (7.98\%) were aware of the W3C Mobile Accessibility Guidelines. Nine participants (4.79\%) were familiar with government guidelines, and one participant was acquainted with the BBC guidelines. Notably, 16 participants (8.51\%) reported being unfamiliar with any accessibility guidelines or standards. The most common two-guideline combination was Android Accessibility Guidelines and Apple iOS Accessibility Guidelines, selected by 35 participants (27.13\%). %The most frequent three-guideline combinations included WCAG, Android Accessibility Guidelines, and Apple iOS Accessibility Guidelines, reported by 14 participants (19.18\%).

Next, we examined guideline familiarity based on the participants’ app development platform (survey question \#9). Among iOS developers, 61 out of 75 (81.33\%) reported being familiar with Apple’s accessibility guidelines. Additionally, 39 iOS developers indicated familiarity with Android’s guidelines, 15 were aware of WCAG, and eight were unfamiliar with any standards. Among Android developers, 62 out of 77 (80.52\%) were familiar with Android accessibility guidelines, while 39 knew of iOS guidelines, 18 were aware of WCAG, and nine were unfamiliar with any standards. Finally, a chi-square test shows a statistically significant association between development platform and guideline familiarity ($\chi^2$(5) = 31.65, p < 0.05), suggesting that the platform a developer specializes in may influence their accessibility knowledge.

\vspace{1mm}
\noindent\textbf{\textit{RQ1 Summary:}}
Developers generally perceive accessibility as important, with 49.09\% rating it as "Very important." However, familiarity with accessibility guidelines is limited, with most developers recognizing platform-specific guidelines like Android (35.11\%) and iOS (32.98\%) accessibility guidelines, while only 10.11\% are aware of universal standards like WCAG. This highlights a gap between perception and knowledge of accessibility practices.

\subsection*{\RQB}
Building upon our RQ1 findings, this RQ examines both implemented accessibility features and integration challenges, aiming to identify actionable insights for addressing implementation barriers. We address this RQ through two sub-RQs.

\textbf{RQ2a: Which accessibility features are frequently implemented in mobile apps?}
To answer this sub-RQ, we analyze survey questions \#13, \#14, and \#15. In question \#13, an MCQ, we asked participants to indicate the types of software for which they had incorporated accessibility features. Among the participants, 100 (76.34\%) reported implementing accessibility for mobile apps, while 18 (13.74\%) did so for web apps, and five (3.82\%) for desktop apps. Additionally, eight participants selected "Other," referring to custom platforms like TV or smartwatch apps. Only 10 participants did not implement accessibility for mobile apps.

\begin{table}[!htbp]
\caption{Top 5 accessibility features adopted by developers.}\vspace{-3mm}
\label{tab: accessibility_feature_adoption}
\small
\begin{tabular}{
|p{0.28\textwidth}|
>{\raggedleft\arraybackslash}p{0.05\textwidth}|
>{\raggedleft\arraybackslash}p{0.05\textwidth}|
}
\hline
\multicolumn{1}{|c|}{\textbf{Answer Option}}                         & \multicolumn{1}{c|}{\textbf{Count}} & \multicolumn{1}{c|}{\textbf{Percentage}}       \\ \hline
Resizable Text and Scalable Fonts     & 65                         & 11.67\%                       \\ \hline
Text-to-Speech (TTS)                  & 60                         & 10.77\%                       \\ \hline
Screen Reader Compatibility           & 53                         & 9.52\%                        \\ \hline
Keyboard Accessibility                & 47                         & 8.44\%                        \\ \hline
Alternative Text for Images           & 46                         & 8.26\%                        \\ \hline
\end{tabular}
\end{table}

Participants who selected mobile apps for question \#13 were presented with questions \#14 and \#15. Question \#14, an MCQ, focused on mobile-specific accessibility features typically incorporated into apps. Table \ref{tab: accessibility_feature_adoption} shows the top five features selected by participants. The most common feature was Resizable Text and Scalable Fonts, with 65 responses (11.67\%), followed by Text-to-Speech (TTS) capabilities at 60 responses (10.77\%), and Screen Reader Compatibility at 53 responses (9.52\%). On average, participants reported incorporating 5.51 accessibility features. The most common two-feature combination was Resizable Text and Scalable Fonts with Screen Reader Compatibility (41 times), while the most frequent triad combination included Resizable Text and Scalable Fonts, Screen Reader Compatibility, and TTS (26 times).% (0.62\%). 

To ensure that we captured all accessibility features used by participants, we included an optional free-text question (survey question \#15). Seven participants responded, highlighting features such as GPS, device sensors (accelerometer, gyroscope, magnetometer, barometer), AI for speech-to-text conversion, and disabling animations to reduce motion discomfort.

There is a statistically significant relationship between developers' accessibility guidelines familiarity and the number of accessibility features they implemented (ANOVA: F = 13.51, p < 0.05; Spearman $\rho$ = 0.39, p < 0.05), suggesting that developers with greater guideline familiarity tend to implement more accessibility features.

\textbf{RQ2b: What obstacles impede accessibility implementation in mobile apps?}
We address this sub-RQ through survey questions \#16 and \#17. 

Participants with experience in mobile app accessibility (survey question \#13) were asked if they faced challenges during integration (question \#16). Out of 110 respondents, 56 (50.91\%) answered "Yes" and 45 (40.91\%) said "No." Those who answered "Yes" provided additional details in a free-text survey question (question \#17). Through thematic analysis, we identified a diverse range of challenges that developers encounter during implementation. From these responses, we derived a \rev{categorization} of accessibility implementation barriers, which groups the challenges across three dimensions: technical, organizational, and ecosystem-level. 

\noindent\textbf{Technical Barriers.}
\begin{itemize}
    \item \textbf{API/Framework/Library Limitations:} 
A common theme noted in 17 responses is the challenges developers face due to limitations in APIs from platform vendors and third-party frameworks. These issues often lead to bugs, incompatibilities, and inadequate support for accessibility features, complicating their implementation. For instance, while cross-platform tools and hybrid frameworks can reduce development time and costs, they frequently lack strong accessibility support. One respondent remarked, ``\textit{Low support for the programming language I use in terms of accessibility}," indicating that the choice of programming language can further limit accessibility capabilities. As a result, developers may need to write extra code to address the missing functionality.

 \item \textbf{Technical Implementation Challenges:} With 16 responses, this theme covers technical difficulties developers encounter when coding and implementing accessibility features. Challenges include making complex UI components accessible, handling dynamic content, ensuring text scalability, using appropriate color schemes, and integrating assistive technologies like screen readers. Refactoring existing code for better accessibility also poses significant challenges, as one respondent noted, ``\textit{adding to the existing codebase often requires restructuring}.''
\end{itemize}

\noindent\textbf{Organizational Barriers.}
\begin{itemize}
    \item \textbf{Organizational/Business Constraints:} This theme centers around non-technical barriers to accessibility implementation due to business priorities, resource allocation, and organizational culture, and is associated with 14 responses. Developers face challenges such as insufficient time allocated to accessibility implementation in project timelines, limited stakeholder and employer interest, and budget constraints. These constraints often lead to accessibility being treated as an afterthought rather than an integral part of the development process, creating frustration for both developers and end users. As one developer noted,  ``\textit{Most of the time the UI and business logic are done without considering the accessibility implementation. So usually, it requires rework}." 
\end{itemize}

\noindent\textbf{Ecosystem-level Barriers.}
\begin{itemize}
    \item \textbf{Documentation Shortcomings:} With 10 responses, this theme relates to inadequate or outdated documentation around incorporating accessibility into apps. This includes documentation that is hard to understand, does not cover all necessary features, outdated solutions, or lacks concrete examples for specific requirements. The absence of comprehensive and up-to-date documentation can lead to reduced productivity for developers; as one developer stated, ``\textit{some documentations are not always properly done hence it takes time figuring out what needs to be done}."

    \item \textbf{Platform and Device Fragmentation:} This theme relates to challenges caused by the need to support various devices and operating systems and is associated with nine responses. The existence of different versions of Android and iOS, along with varying hardware capabilities and features, makes it challenging to implement a uniformly accessible solution. As one developer noted, ``\textit{Multiple versions of Android behave differently. Some were working some were not}," emphasizing the compatibility issues.% across devices.

    \item \textbf{Accessibility For All/Inclusive Design:} This theme relates to the challenges of designing apps that work effectively for users with diverse disabilities. As one respondent admitted, ``\textit{It's difficult to tell beforehand what accessibility features your users need the most}." Developers struggle to understand and address the varied requirements of different disability types and to create intuitive experiences for all users, regardless of ability. This theme was associated with seven responses.
\end{itemize}

\noindent We also identified less common themes (with three or fewer responses), like knowledge gaps, where developers complain about their lack of understanding of accessibility features, and issues in testing accessibility features across various scenarios and devices.

\vspace{1mm}
\noindent\textbf{\textit{RQ2 Summary:}}
Developers primarily implement text-related accessibility features, such as resizable text, text-to-speech, and screen reader compatibility. Challenges include API limitations, technical difficulties, organizational constraints, outdated documentation, and platform fragmentation. Developers with greater familiarity with guidelines tend to implement more features.

\subsection*{\RQC}
While RQ2 examines feature implementation, RQ3 explores how developers verify that these features work properly for users through testing. The findings will help identify potential gaps in testing methodologies that might contribute to accessibility barriers. We address this RQ through three sub-RQs.

\textbf{RQ3a: What are the typical user-reported accessibility issues?}
As part of investigating accessibility testing practices, it is important to understand the real-world challenges that testing needs to address. One effective method for identifying these issues is through end-user feedback. Thus, survey question \#18, an open-ended question, asks participants about accessibility-related feedback they received for their apps. A thematic analysis of the responses yields the following:

\noindent\textbf{Vision Accessibility Issues:} This theme highlights the challenges faced by users with visual impairments while using a mobile app, based on nine responses. Feedback includes issues with screen readers not properly announcing interactive elements, insufficient contrast ratios, inadequate text sizing, and missing alternative text descriptions. For instance, a respondent noted, ``\textit{Some users with vision disabilities report to us problems with Screen Reading - usually it is unclear UI elements description}." Accurate descriptions are crucial for users relying on assistive technologies to navigate. %Users with vision impairments rely on properly implemented descriptions to effectively navigate using assistive technologies. 

\noindent\textbf{Hearing Accessibility Issues:} This theme includes issues affecting end-users who are deaf or hard of hearing. Despite being critical for many users, hearing accessibility received minimal detailed feedback on this question, with only two responses. One respondent noted ``\textit{Users unable to use the application due to deaf or low hearing issues}" without elaborating further. Additionally, one respondent mentioned ``\textit{We received missing accessibility features in audio book app, then implement TTS.}" While text-to-speech benefits individuals with visual impairment, it can also benefit users with hearing impairments when paired with visual cues such as captions.

\noindent\textbf{Motor Control Accessibility Issues:} This theme addresses feedback from end-users with mobility challenges and is associated with two responses. Individuals with motor impairments may struggle with precise tapping, complex gestures like pinch-to-zoom, or interfaces that require quick responses. As one respondent noted,  ``\textit{Users unable to access some function due to mobility control}.'' 

\noindent\textbf{Platform-Specific Issues:} There were three responses that did not touch on a specific disability but were challenges developers face with the platform when incorporating accessibility features. As one respondent mentioned, ``\textit{we need special accessibility but, because of security reasons, apple don't allow us to publish application and we need to clarify the usage for them}", showing how technical and regulatory limitations can hinder accessibility implementation.

\noindent\textbf{Positive User Feedback:} One response highlighted positive feedback: ``\textit{I did receive an email from a blind user who mentioned how grateful he was to have voiceover functionality in one of the apps}." This response shows that incorporating accessibility into a mobile app can genuinely improve the experience for users with disabilities.

\textbf{RQ3b: What practices do developers utilize for accessibility testing?}
For this sub-RQ, we analyze the responses to survey questions \#19 and \#20. 

Through an open-ended question, survey question \#19 asked participants at what stage in the Software Development Life Cycle (SDLC) they typically test the accessibility compliance of their apps. Our analysis provides the following insights\footnote{Nine responses were discarded for being incomplete or irrelevant.}:

\noindent\textbf{Testing Frequently Occurs in Mid-to-Late SDLC Phases:} The findings indicate that most accessibility testing occurs primarily during the Development (28 responses) and Testing (31 responses) phases. Further, within the Testing phase responses, we observe the mention of specialized testers handling accessibility verification. Responses such as ``\textit{Testers do it}," ``\textit{QA testing}," and ``\textit{conducted by manual QA-engineer}" indicate that many development teams delegate accessibility testing to dedicated quality assurance personnel. Additionally, many participants referred to testing occurring ``\textit{after development}" and ``\textit{before release}," which we categorize under End of Development/Pre-release activities, accounting for 18 responses. This pattern suggests that accessibility is often treated as a feature to be implemented and verified rather than a core design consideration from project inception.

\noindent\textbf{Limited Early-Phase and End-to-End Accessibility Accessibility Testing:} Our analysis reveals a considerable gap in both early-phase consideration and continuous testing for accessibility throughout the SDLC. Nine participants indicated that they prioritized accessibility from the start of the project, such as assessing technical feasibility and conducting requirements and design reviews (e.g.,``\textit{checking mockups for accessibility}"). Further, only six participants explicitly stated they test for accessibility throughout the entire SDLC ("\textit{all phases}" or ``\textit{at every stage}"). This highlights a disconnect between current and best practices, which recommend accessibility consideration throughout the SDLC \cite{Bi2022}. \rev{Moreover, many organizations use specialized teams during the Quality Assurance (QA) phase. While this approach is not necessarily problematic, it often leads to accessibility being addressed only during this stage. In agile methodologies, such as Scrum, QA activities typically occur at the end of each sprint. To better integrate accessibility, developers should focus on it from the start of the sprint, ensuring it is part of the development process, rather than an afterthought.}

\noindent\textbf{Post-Deployment Testing:} Four participants indicated they perform accessibility testing post-deployment (e.g., ``\textit{production}" and ``\textit{normally at the time of deployment}"). This reactive approach may identify defects but risks harming the experience for disabled users and could result in negative app reviews.%This post-deployment approach represents a reactive rather than a proactive approach to accessibility compliance. While this approach can identify defects, it may harm the user experience for disabled end-users and could lead to negative app reviews.

\noindent\textbf{Absence of Testing.} Seven mentioned they do not test for accessibility, with one stating, ``\textit{We don't really, I even worked for one of the biggest companies in America and it was never really brought up}." %or ``\textit{No, I don't test them, I will add them, and they should be lucky, if they work}."   

Moving on, survey question \#20 focused on techniques for assessing accessibility compliance in mobile apps. Responses showed that 88 participants (43.14\%) use manual testing, 46 (22.55\%) use automated testing, and 27 (13.24\%) leverage assistive technologies (e.g., screen readers and switches) during testing. Additionally, 20 participants (9.80\%) conducted user testing with individuals with disabilities, while two responses (0.98\%) fell under ``Other," and included automated testing via unit tests and the XCTest framework.
Finally, we also found that developers with higher accessibility guideline familiarity used more testing techniques (ANOVA: F = 15.37, p < 0.05; Spearman $\rho$ = 0.44, p < 0.05), showing that knowing the standards is important for effective testing practices.

\textbf{RQ3c: What resources would improve accessibility testing?}
This sub-RQ aims to identify enhancements to support mobile app developers in implementing accessibility standards. Specifically, survey question \#21 addresses software technology improvements that developers desire, while question \#22 explores the resources developers use. Together, they offer insights into better supporting developers in adding accessibility features to their apps.

A thematic analysis of the free-text responses to question \#21 yields the following areas:\footnote{Seven responses were discarded for being incomplete or irrelevant, and three mentioned no improvements.}

\noindent\textbf{Developer Tooling \& Platform Support:} This theme focuses on enhancing technical infrastructure for better accessibility implementation, gathered from 12 responses. Focus areas include improved integration of accessibility tools in development environments like Android Studio for real-time feedback, cross-device resources, platform APIs for quicker implementation, and affordable automated testing tools, especially open-source options. Additionally, user simulation tools are needed to help developers understand the experiences of users with impairments.

\noindent\textbf{Tool Usability:} With six responses, this theme focuses on usability improvements to tools and technologies related to accessibility implementation. Our findings highlight the need for simpler implementation processes that avoid complex code changes, provide predictable outcomes, and minimize the learning curve.

\noindent\textbf{AI/ML Solutions:} This theme is associated with four responses and focuses on using AI/ML techniques, such as large language models (LLMs), to assist with accessibility implementation through code suggestions and automation, including the identification of elements missing proper accessibility attributes.

\noindent\textbf{Knowledge Resources:} With four responses, this theme highlights the need for accessible, well-written documentation to support accessibility implementation. Participants emphasize the importance of clear explanations, best practices, and realistic examples.

\begin{table}[!hbtp]
\centering
\caption{Resources utilized by app developer for accessibility.}
\label{Table:Resources}\vspace{-3mm}
\small
\resizebox{\columnwidth}{!}{%
\begin{tabular}{
@{}
>{\raggedright\arraybackslash}p{0.36\linewidth}
p{0.47\linewidth}
>{\raggedleft\arraybackslash}p{0.10\linewidth}
@{}
}
\toprule
\multicolumn{1}{c}{\textbf{Resource Type}} & \multicolumn{1}{c}{\textbf{Example}} & \multicolumn{1}{c}{\textbf{Count}}\\ \hline
Official Platform Documentation & Google/Apple documentation or guidelines        & 58 \\ \hline 
Online Articles and Blogs       & Articles/Blogs hosted on Medium              & 21 \\ \hline
Online Video Tutorials          & YouTube video tutorials                      & 13 \\ \hline
Community Knowledge             & Stack Overflow, seniror developers           & 11 \\ \hline
Formal Learning Resources       & Books, attending conferences, online courses & 8  \\ \hline
AI Chatbot                      & ChatGPT                                      & 6  \\ %\midrule
\hline
\end{tabular}%
}
\end{table}

Moving on to question \#22, Table \ref{Table:Resources} shows our analysis of the free-text responses regarding the resources participants turn to for assistance with mobile app accessibility implementation.  
Developers utilize a variety of resources for assistance with incorporating accessibility into their apps, with 58 participants highlighting official platform documentation from Google and Apple. They also utilize online articles and blogs, especially from platforms like Medium, for practical insights and best practices. Other resources include video tutorials on YouTube for hands-on demonstrations, community knowledge forums such as GitHub, Stack Overflow, and Reddit, and also requesting assistance from senior developers. Additionally, some opt for formal learning through books and professional events. Further, generative AI (genAI) tools like ChatGPT are increasingly influencing how developers access accessibility information.%developers prefer formal learning through books and professional events for a more structured approach. The emergence of generative AI tools, like ChatGPT, is also influencing how developers access accessibility information. %Lastly, there were also responses that were irrelevant or too vague for interpretation.

\vspace{1mm}
\noindent\textbf{\textit{RQ3 Summary:}}
Accessibility testing usually happens in mid-to-late development phases, with limited early-phase and continuous testing. While developers use manual testing, automated tools, and assistive technologies, only 9.8\% involve users with disabilities in the process. Developers who tend to be more familiar with accessibility guidelines use various testing techniques. End-users often report vision-related issues. There's a need for improved tools, better documentation, and AI/ML solutions to improve testing practices.

\subsection*{\RQD}
This RQ examines ten participants who reported not implementing accessibility features in their mobile apps (via survey question \#13). It aims to identify barriers to adoption and explore changes that could encourage these practices. We analyze open-ended responses from survey questions \#23 and \#24 to address this RQ.

\noindent\textbf{\textit{Barriers.}} Responses to survey question \#23 reveal following themes:

\noindent\textbf{Lack of Requirements}: Four participants cited the lack of accessibility requirements as a barrier to implementing accessibility features. Responses like ``\textit{No requirement from product team,"} highlight the absence of top-down mandates for accessibility integration.

\noindent\textbf{Limited Experience}: Three participants cited insufficient experience with mobile development or accessibility practices as a barrier. As one developer noted, ``\textit{I do not really have much experience developing mobile apps, so there was no need for accessibility features."}

\noindent\textbf{Feature Prioritization}: Two responses reflected a tendency to deprioritize accessibility in favor of business-critical functionality. One participant explained, ``\textit{All the apps I worked on were less popular, so they were always on the MVP stage. This is the reason that we did not implement those features."}

\noindent\textbf{\textit{Enablers.}} Examining the responses to survey question \#24, yields the following themes:

\noindent\textbf{Formal Requirements}: Six participants expressed that formal inclusion of accessibility in product requirements or development tickets would significantly influence adoption. Furthermore, understanding the accessibility needs of the target users is essential for incorporating accessibility, as it can lead to innovative solutions and improved user satisfaction. Additionally, while accessibility features benefit users with disabilities, they should also ensure a return on investment for developers, as noted by one respondent:  ``\textit{The audience of the app needs these features explicitly, or the app must give a stable income that allows it to add this feature.}''

\noindent\textbf{Professional Development}: Three participants noted the need for additional technical expertise or hands-on exposure. For example, one respondent stated, ``\textit{Gain more experience with mobile app development and create an app that would make sense to have accessibility features."}

\vspace{1mm}
\noindent\textbf{\textit{RQ4 Summary:}}
Non-adopters of accessibility features face three main barriers: lack of formal requirements, limited experience, and prioritizing core functionality over accessibility. To improve adoption, accessibility should be formally included in product requirements, along with improving developer skills.

\subsection*{\RQE}
This RQ examines how accessibility practices differ based on the platform developers specialize in (iOS vs. Android) and their level of experience. By analyzing survey responses, we identify trends in feature implementation, testing practices, and challenges faced by developers across these dimensions. 

\noindent\textbf{Platform Specialization.}
iOS developers faced challenges like framework limitations (11 responses) and technical implementation issues (11) related to accessibility. Other barriers included organizational constraints (9) and documentation gaps (6). Accessibility testing mainly occurred in later phases: during development (29), QA (22), and before release (11). Most used manual inspections (61) and automated tools (33), while seven did no formal testing.

Android developers reported similar challenges, primarily due to framework or library limitations (14 responses), organizational constraints (12), and technical implementation issues (10). Only one developer cited knowledge gaps, indicating better community support compared to iOS. In testing practices, 30 developers conducted accessibility testing during development, 21 during testing, and 12 pre-release. Manual testing (65) and automated tools (36) were commonly used, with ten developers not conducting any accessibility testing, highlighting a need for ongoing support and training.

\noindent\textbf{Developer Experience.}
We grouped developers into three categories based on self-reported years of app development: Junior: $\leq$ 2 years, Mid-level: 3–5 years, and Senior: $\geq$ 6 years, aligning with common industry stages \cite{indeed,dept}.% This grouping aligns with common professional development stages used in industry \cite{indeed,dept}. 

\noindent\textit{\textbf{Junior Developers:}}  
Among the 15 junior developers, six were unaware of mobile accessibility guidelines, and only two were familiar with multiple guidelines. Most did not identify any accessibility-related challenges, likely reflecting limited awareness rather than the absence of issues. When challenges were mentioned, four developers indicated third-party library constraints and a lack of guidance on accessible design patterns. Accessibility testing typically occurred late in the development cycle, primarily during QA and pre-release for six developers and post-release for one. They mostly relied on manual testing, with little use of automation. Regarding feature implementation, they commonly added text scaling and high-contrast UI options, which usually require little configuration.%Testing of accessibility features typically occurred late in the development cycle, primarily during QA for eight developers and post-release for three developers. Manual testing was the most common method used, with minimal reliance on automation or proactive evaluation.

\noindent\textit{\textbf{Mid-level Developers:}}
Out of 47 mid-level developers, 24 reported familiarity with two or more guidelines. The challenges they faced were more diverse: seven developers mentioned limitations with third-party libraries, six pointed to unclear documentation, and five noted organizational deprioritization. Testing practices were better integrated, with 18 developers testing during development, 17 during QA, and seven during pre-release. However, early-stage testing was limited; only four developers included accessibility testing throughout the SDLC, and three considered accessibility in the planning phase.
This group implemented a broader set of features than junior developers, frequently including screen reader compatibility, dynamic text sizing, and contrast settings. Some developers also reported support for color-blindness and custom label assignment. Overall, their feature coverage indicates growing maturity in accessibility implementation, but occasional gaps.

\noindent\textit{\textbf{Senior Developers:}}
Among 48 senior developers, 19 were familiar with multiple accessibility standards, while three had no familiarity. The challenges faced by these developers shifted from technical issues to systemic and architectural concerns. Specifically, 11 developers highlighted complex UI flows as a challenge, while nine pointed to business constraints such as tight deadlines or limited stakeholder support. Additionally, five developers noted platform fragmentation issues. Testing was more comprehensively integrated across the SDLC: 21 tested during development, 12 during QA, and six during planning. Two reported testing at every stage. The tools used were diverse, suggesting awareness of multiple techniques: 40 developers performed manual testing, 18 utilized automation tools, 15 employed assistive technologies, and 10 gathered feedback from users.
They also implemented a wide range of accessibility features, including screen reader integration, keyboard navigation, using custom accessibility labels, support for physical input devices, reduced motion modes, and assistive technology API compatibility. This approach reflects a deeper understanding of accessibility and a focus on creating a user-friendly experience for all.

\noindent\textbf{Statistical Analysis:} 
Finally, a one-way ANOVA conducted on the number of accessibility features implemented across experience groups yields a marginally significant result: F(2, N) = 3.03, p = 0.052. 
Although the result is marginally above the conventional 0.05 threshold, the trends indicate that senior developers engage more with accessibility features, suggesting that experience plays an important role in influencing accessibility maturity.

\vspace{1mm}
\noindent\textbf{\textit{RQ5 Summary:}}
Accessibility practices improve with experience. Junior developers focus on basic features and late-stage testing, mid-level developers implement broader features but face diverse challenges, and senior developers integrate accessibility across the SDLC, using advanced tools and techniques.

\section{Related Work}
\label{Section:related}

\subsection{Repository Mining}
Chen et al. \cite{chen2021accessible} developed Xbot, a UI exploration tool that outperforms Google Monkey in detecting accessibility issues. Analyzing 2,270 Android apps, Xbot found 86,767 defects. Key violations include small touch targets, low contrast, and missing semantic labels. Yan et al. \cite{yan2019current} utilized the IBM Mobile Accessibility Checker to evaluate 479 Android apps, finding a violation rate of 94.8\% and identifying six core barriers to accessibility. They propose metrics like the Inaccessible Element Rate and advocate for integrating automated accessibility scanners into CI/CD pipelines to enforce compliance cost-effectively. Oliveira et al. \cite{oliveira2024exploring_1} explored user reviews of mobile apps, revealing accessibility issues related to improper color schemes, small text, and missing labels, which lead to UI misinterpretation and social exclusion. %They recommended enhancements such as colored labels, dynamic theming, and better screen-reader compatibility, while also highlighting the negative experiences of disabled users, including navigational challenges and feelings of discrimination. 
Ross et al. \cite{ross2020epidemiology} analyzed accessibility barriers in over 9,000 free Android apps, finding that missing labels and undersized elements were the most common issues, especially in Image Buttons and Clickable Images. They noted severe barriers like the scarcity of TalkBack-focusable elements, making apps unusable for assistive technologies. Additionally, apps developed with hybrid tools (e.g., Adobe Air, Unity) faced more accessibility challenges. %The authors recommend educating developers on accessibility, enhancing tools with automated checks, improving guidelines, and promoting inclusive design practices.
Ballantyne et al. \cite{Ballantyne2018} conducted a heuristic evaluation of 25 popular Android applications to assess their adherence to mobile-specific accessibility guidelines, revealing widespread accessibility violations at the design and content levels and highlighting systemic issues in app development practices. % along with the need for improved accessibility awareness among developers.

Vendome et al. \cite{vendome2019can} conducted a mining analysis of over 13,000 Android apps and a qualitative study of 366 Stack Overflow discussions on accessibility. They found only 2.08\% of apps utilized accessibility APIs, while 50.08\% included assistive content descriptions for all GUI elements. Their analysis of developer discussions shows that discussions primarily focused on supporting visually impaired users.
Similarly, Indika et al. \cite{indika2024exploring} reviewed 15 years of mobile accessibility discussions, analyzing 3,022 questions that highlighted challenges like integrating screen readers and ensuring accessible UI elements.
Ma et al. \cite{ma2022first} studied dark mode implementation by examining 324 Stack Overflow posts and over 6,000 apps, identifying issues such as widget design and implementation errors.

Mateus et al. \cite{Mateus2021} studied accessibility issues for visually impaired users on websites and mobile apps. They found that automated tools identified less than 40\% of web issues and under 20\% of mobile app issues. Expert inspections offered broader insights but missed user-specific barriers. User evaluations were most effective in pinpointing critical problems, such as inconsistent content organization and poor navigation, affecting user experience.
Olivera et al. \cite{oliveira2024exploring_2} conducted a systematic review analyzing mobile app user reviews for accessibility violations. Despite advancements in mobile accessibility, issues like poor color contrast, illegible fonts, inconsistent navigation, and limited customization persist.
Acosta et al. \cite{acosta2021accessibility} conducted a scoping review on accessibility tools, WCAG guidelines, and design methods for mobile accessibility, highlighting a focus on sensory impairments, followed by cognitive and motor disabilities, and favoring manual expert testing over automated. % Sensory impairments dominate the literature (86\%), followed by cognitive (9\%) and motor (5\%) disabilities. Manual, expert‑led accessibility testing is most prevalent (50\%), compared to automated (18.2\%) and heuristic evaluations (18.2\%). %The authors endorse WCAG as the de facto framework and call for integrating automated and heuristic testing tools into agile development lifecycles, alongside strengthened regulations to mandate participatory design with disabled users.

\subsection{Practitioner Surveys}
Alshayban et al. \cite{alshayban2020accessibility} conducted a large-scale empirical analysis of over 1,000 Android apps, surveyed 61 developers, and analyzed user reviews to understand accessibility challenges and their impact. The study found that accessibility issues are widespread in Android apps, with common problems like text contrast, touch target size, and image contrast. Additionally, developers often lack awareness of accessibility principles, and organizations deprioritize accessibility, leading to persistent issues. %Further, while users complain about missing labels, text size/color, and image/icon contrast, these issues do not significantly impact app ratings or popularity.
Similarly, Di Gregorio et al. \cite{di2022making} conducted an empirical study where they manually tested 50 top-rated Android apps and surveyed 75 developers to understand accessibility practices in Android mobile app development. The findings show a low implementation of accessibility guidelines, with only 41\% of guidelines applied to the analyzed apps. Developers cited challenges such as personal choices, company policies, vague standards, and socio-technical barriers to implementing guidelines. Additionally, the study highlighted the lack of automated tools and standardized methods to support accessibility implementation.
Leite et al. \cite{leite2021accessibility} found that many mobile app developers in Brazil are unaware of accessibility issues and face barriers like insufficient training, time constraints, and a lack of requirements.

\rev{
\subsection{Comparison with Prior Studies}
Previous work on mobile app accessibility includes mining studies and developer surveys. Mining studies revealed widespread accessibility problems but offered little insight into the \textit{human or organizational factors} behind them. Survey-based studies provided such context but were largely Android-only or regionally limited. Our work provides an updated perspective by providing a global, cross-platform, mixed-method investigation of accessibility practices among mobile developers. Below, we highlight how our findings confirm and extend upon previous survey-based studies.
}

\noindent\rev{\textbf{Awareness, Knowledge, and Perceived Importance.} Our findings align with prior research, indicating that while developers recognize the importance of accessibility, their knowledge of guidelines is limited (\cite{di2022making,leite2021accessibility,alshayban2020accessibility}). Our study extends this evidence by quantifying the attitude-practice gap between perceived importance and actual implementation. It also identifies platform-specific awareness differences between Android and iOS and provides statistical evidence linking guideline familiarity to both platform specialization and developer experience relationships.
}

\noindent\rev{\textbf{Implementation of Accessibility Features.} Prior work either focused on Android features (\cite{di2022making}) or assistive technology familiarity (\cite{leite2021accessibility}), showing that developers mainly focus on text and vision accessibility. Our study provides a cross-platform quantitative analysis of implemented features, showing that developers with higher familiarity with guidelines implement significantly more features.
}

\noindent\rev{\textbf{Testing Practices.} Prior studies reported late and mostly manual testing (\cite{di2022making,leite2021accessibility}).
We extend these findings by providing a cross-platform breakdown of testing across SDLC phases, which shows that testing diversity increases with developer experience.
}

\noindent\rev{\textbf{Tooling.} Confirms prior work that tool use remains limited (\cite{di2022making,alshayban2020accessibility}). Our findings highlight an emerging reliance on AI-based tools and detail the specific improvements that developers seek.
}

\noindent\rev{\textbf{Barriers and Challenges.} We confirm prior findings on obstacles such as time, training, and tool gaps (\cite{di2022making,leite2021accessibility,alshayban2020accessibility}). We introduce a structured, multi-level categorization that clarifies why barriers persist, and propose an ecosystem-level dimension of challenges.
}

\noindent\rev{\textbf{Cross-Platform and Global Scope.} Unlike previous studies that focused on a single platform (\cite{di2022making,alshayban2020accessibility}) or region (\cite{leite2021accessibility}), our research spans across 43 countries and covers both Android and iOS. This broader view exposes shared and distinct accessibility practices, challenges, and contextual factors.
}

\section{Discussion and Implications}
\label{Section:discussion}
\rev{This study reveals that accessibility practices are shaped by platform ecosystems, developer experience, and organizational contexts. By examining iOS and Android developers, we not only identify shared practices but also uncover how platform-specific constraints influence accessibility, providing insights not possible from single-platform studies. For instance, iOS developers report framework limitations due to Apple's closed ecosystem and strict API controls, while Android developers struggle with fragmentation across device manufacturers and OS versions.}

\noindent\textbf{The Platform-Practice Gap.}
While nearly 93\% of respondents view accessibility as ``moderately important,'' our findings reveal a limited implementation depth. While developers adopt a broad range of accessibility features, most of the features they implement primarily address visual impairments. Developers focus on a narrow set of text-related features (e.g., resizable fonts, TTS), with little attention to supporting users with motor, cognitive, or hearing impairments. \rev{This imbalance suggests that developers' understanding of accessibility remains narrow, constrained by familiar use cases and limited exposure to diverse user needs.} Furthermore, while platform-specific guidelines are well-known, only a minority of participants were aware of platform-independent frameworks, such as WCAG or W3C guidelines. This lack of awareness highlights a gap in understanding, which results in implementations that often fall short of being comprehensive or inclusive in practice.
 
\noindent\textbf{Experience Matters, But It’s Not Enough.}
We observed a correlation between professional experience and accessibility practices. Junior developers often rely on default settings or basic features, such as text scaling, while senior developers are more likely to implement comprehensive support for screen readers and external input devices. This suggests that knowledge of accessibility is acquired gradually through experience rather than through structured onboarding or formal education. Senior developers tend to utilize more sophisticated testing techniques, including assistive technologies and user feedback loops, yet accessibility is often integrated late in the development process. This highlights that without systemic changes, like integrating accessibility into the planning stages, creating specific standards, and improving training, accessibility will remain a reactive and inconsistent practice.

\noindent\textbf{\rev{Multi-Dimensional Accessibility Barriers.}}
\rev{Our findings show that accessibility challenges in mobile app development extend beyond isolated technical issues and reflect the interaction of multiple socio-technical layers.
Although our qualitative data do not reach theoretical saturation, they provide a structured, cross-platform view of how technical, organizational, and ecosystem factors constrain accessibility work. While confirming prior work about time pressure, limited training, and tools, our analysis adds an ecosystem-level dimension that highlights inconsistencies in documentation, fragmented standards, and platform-specific issues. This categorization offers a structured understanding of the factors that hinder accessibility adoption, providing a means for targeted interventions by tool builders, educators, researchers, and platform vendors.}

\noindent\rev{\textbf{Accessibility Debt.}
Our findings also contribute to the concept of accessibility debt, a type of technical debt \cite{ernst2021} caused by the accumulation of accessibility issues resulting from delayed or inconsistent integration of accessibility practices throughout the SDLC \cite{sklavenitis2025scoping, churchill2018putting}. Developers in our study frequently reported that accessibility was addressed only at late testing stages, if at all. This deferred attention can lead to rework, increased costs to fix problems, and a poor user experience \cite{suvvari2023shift}. Unlike traditional technical debt, which primarily impacts code maintainability, accessibility debt has a direct impact on individuals and is influenced by socio-technical factors such as organizational priorities and platform constraints. Addressing it requires integrating inclusive design principles across the SDLC, rather than treating accessibility as a post-release fix.
}

\noindent\textbf{Implications for Research, Practice, and Education.}
\begin{itemize}
\item \rev{\textbf{Educators and Trainers.} Accessibility needs to move beyond awareness campaigns and into hands-on curriculum and developer upskilling programs. Many developers are self-taught %(56.36\%) 
and may lack familiarity with universal standards, such as WCAG. %(10.11\%)
Educational initiatives must prioritize accessibility as a fundamental skill in app development, integrating it into project-based learning rather than treating it as an afterthought. To achieve this, educators and trainers should promote universal design principles, making accessibility a core component of software engineering education through hands-on projects and collaboration with subject-matter experts to improve content relevance.}

\item \rev{\textbf{Tool and Platform Vendors.} Developers encounter challenges such as fragmented APIs and poor integration of feedback in IDEs, making it essential for tool and platform vendors to prioritize real-time accessibility feedback alongside more traditional concerns of performance and security. By providing code suggestions and automated diagnostics, vendors can improve inclusive design. Additionally, offering simulation tools that preview app behavior with respect to various impairments allows for early and more comprehensive identification of accessibility issues. Implementing real-time, context-aware feedback within IDEs and CI pipelines, along with AI-assisted tools that simulate impairments during the design phase, can reduce reliance on late-stage QA and ensure a more accessible user experience.}

\item \rev{\textbf{Developers and Organizations.} Accessibility should be prioritized throughout the SDLC, not just during QA. Developers must go beyond minimal compliance by proactively learning about cross-disability needs and incorporating testing earlier in the process. Organizations should embed accessibility as a core quality attribute, recognizing it in team goals and success metrics and encouraging an accessibility-first culture by integrating it into planning, requirements, and definition-of-done criteria, along with measurable outcomes. This can be supported through training, mentorship, and dedicated time to address accessibility debt. Promoting cross-functional collaboration among design, development, and QA teams will ensure that accessibility remains a shared responsibility, leading to improved adoption and contributions to open-source accessibility tools.}

\item \rev{\textbf{Researchers.} The gap between perceived importance and actual adoption of accessibility features suggests the concept of accessibility debt, a form of technical debt. This debt accumulates when organizations neglect to implement necessary accessibility measures, leading to long-term effects that impact both users and organizations. Future research could explore the concept of accessibility debt in greater detail, examining how it accumulates, how it differs from technical debt, its long-term repercussions, and potential interventions to reduce this debt.}

\end{itemize}

\section{Threats To Validity}
\label{Section:threats}
This study's validity is impacted by several factors. Self-selection bias is one concern, as developers with a stronger interest in accessibility may have been more likely to participate. Despite reaching participants from 43 countries, the uneven geographic distribution limits the generalizability of our findings. While platforms like Reddit and GitHub were options, LinkedIn provides more detailed user profiles and search capabilities. However, it is not always possible to verify the authenticity of these user profiles, a common challenge in survey-based studies. Additionally, self-reported data may introduce reporting bias, as participants may potentially exaggerate or downplay their experiences. \rev{Since survey-based findings often reflect self-reported practices, they may not accurately represent what developers actually do in practice.} To promote honest feedback, we ensured the survey was anonymous and non-compensated. \rev{Furthermore, classifying developers by self-reported years of experience may not accurately indicate competency, as a junior developer in an accessibility-focused organization might outperform a senior developer without such experience.} Although the use of single- and multiple-choice questions in our survey may seem restrictive in allowing participants to express their views, the predefined answer options were determined based on existing literature. We also included an ``Other'' free-text field in some questions to capture additional perspectives. To address subjectivity in open-ended responses, two authors independently coded the answers and collaborated to refine the themes. \rev{However, some free-text questions had limited and brief responses, making these insights exploratory rather than exhaustive}. %\amila{Furthermore, there are imbalances in the free-text responses grouped by years of experience and platform. These different sample sizes may result in non-representative and less diverse groups, which can affect the insights drawn from free-text analysis}. 
Finally, due to the survey's anonymity, we were unable to seek further clarification or insights.  

\section{Conclusion \& Future Work}
\label{Section:conclusion}
As mobile apps become increasingly essential for accessing information and services, developers must ensure that their apps are usable by people of all abilities, not only because of social responsibilities but also because of legal requirements.
Our survey of 110 developers across 43 countries revealed that while most recognize accessibility's importance, implementation remains challenging. Developers primarily focus on text-related features while struggling with API limitations and organizational constraints, among others. Further, testing typically occurs late in development, with primary barriers including a lack of formal requirements and limited expertise.
Our future work includes longitudinal case studies with development teams to identify best practices and strategies for integrating accessibility in the app development lifecycle.

\bibliographystyle{ACM-Reference-Format}
\balance
\bibliography{main}

@article{Armstrong1997,
  title = {The Place of Inter-Rater Reliability in Qualitative Research: An Empirical Study},
  volume = {31},
  ISSN = {1469-8684},
  url = {http://dx.doi.org/10.1177/0038038597031003015},
  DOI = {10.1177/0038038597031003015},
  number = {3},
  journal = {Sociology},
  publisher = {SAGE Publications},
  author = {Armstrong,  David and Gosling,  Ann and Weinman,  John and Marteau,  Theresa},
  year = {1997},
  month = aug,
  pages = {597–606}
}

@article{Campbell2013,
  title = {Coding In-depth Semistructured Interviews: Problems of Unitization and Intercoder Reliability and Agreement},
  volume = {42},
  ISSN = {1552-8294},
  url = {http://dx.doi.org/10.1177/0049124113500475},
  DOI = {10.1177/0049124113500475},
  number = {3},
  journal = {Sociological Methods\&; Research},
  publisher = {SAGE Publications},
  author = {Campbell,  John L. and Quincy,  Charles and Osserman,  Jordan and Pedersen,  Ove K.},
  year = {2013},
  month = aug,
  pages = {294–320}
}

@article{Bi2022,
  title = {Accessibility in Software Practice: A Practitioner’s Perspective},
  volume = {31},
  ISSN = {1557-7392},
  url = {http://dx.doi.org/10.1145/3503508},
  DOI = {10.1145/3503508},
  number = {4},
  journal = {ACM Transactions on Software Engineering and Methodology},
  publisher = {Association for Computing Machinery (ACM)},
  author = {Bi,  Tingting and Xia,  Xin and Lo,  David and Grundy,  John and Zimmermann,  Thomas and Ford,  Denae},
  year = {2022},
  month = jul,
  pages = {1–26}
}

@misc{linkedinPolicy,
author = {{Linkedin}},
title = {Professional community policies},
url = "https://www.linkedin.com/legal/professional-community-policies",
month = jul,
year = {2025},
note = "Accessed: 2025-7-7"
}

@misc{indeed,
author = {{Indeed}},
title = {Understanding the 10 Career Levels for Software Engineers},
url = "https://www.indeed.com/career-advice/finding-a-job/engineer-level",
month = jun,
year = {2025},
note = "Accessed: 2025-7-7"
}

@misc{dept,
author = {{Dept}},
title = {Junior vs. Mid vs. Senior software engineers – experience, skills, \& expectations},
url = "https://www.deptagency.com/insight/junior-vs-mid-vs-senior-software-engineers-experience-skills-expectations/?utm_source=chatgpt.com",
month = apr,
year = {2021},
note = "Accessed: 2025-7-7"
}

@article{Mateus2021,
  title = {A Systematic Mapping of Accessibility Problems Encountered on Websites and Mobile Apps: A Comparison Between Automated Tests,  Manual Inspections and User Evaluations},
  volume = {12},
  ISSN = {2763-7719},
  url = {http://dx.doi.org/10.5753/jis.2021.1778},
  DOI = {10.5753/jis.2021.1778},
  number = {1},
  journal = {Journal on Interactive Systems},
  publisher = {Sociedade Brasileira de Computacao - SB},
  author = {Mateus,  Delvani Ant\^onio and Silva,  Carlos Alberto and De Oliveira,  Arthur F. B. A. and Costa,  Heitor and Freire,  André Pimenta},
  year = {2021},
  month = nov,
  pages = {145–171}
}

@misc{guidelines-wcag,
author = {{World Wide Web Consortium}},
title = {W3C Accessibility Standards Overview},
url = "https://www.w3.org/WAI/standards-guidelines/",
month = feb,
year = {2024},
note = "Accessed: 2025-4-3"
}

@misc{guidelines-EN01549,
author = {ETSI},
title = {EN 301 549},
url = "https://www.etsi.org/deliver/etsi_en/301500_301599/301549/03.02.01_60/en_301549v030201p.pdf",
month = mar,
year = {2021},
note = "Accessed: 2025-4-3"
}

@misc{guidelines-ISO9241171,
author = {{International Organization for Standardization}},
title = {ISO 9241-171:2008},
url = "https://www.iso.org/standard/39080.html",
month = jul,
year = {2008},
note = "Accessed: 2025-4-3",
year = {2025},
}

@misc{guidelines_android,
author = {Google},
title = {Build accessible apps},
url = "https://developer.android.com/guide/topics/ui/accessibility",
note = "Accessed: 2025-4-3",
year = {2025},
}

@misc{guidelines_ios,
author = {Apple},
title = {Building accessible apps},
url = "https://developer.apple.com/accessibility/",
note = "Accessed: 2025-4-3",
year = {2025},
}

@misc{guidelines_ios2,
author = {Apple},
title = {Accessibility Design},
url = "https://developer.apple.com/design/human-interface-guidelines/accessibility",
note = "Accessed: 2025-4-3",
year = {2025},
}

@misc{guidelines_android2,
author = {Google},
title = {Accessibility Design},
url = "https://m3.material.io/foundations/overview/principles",
note = "Accessed: 2025-4-3"
}

@misc{web-appStoreCategory,
author = {Statista},
title = {Apple: most popular app store categories 2024},
url = "https://www.statista.com/statistics/270291/popular-categories-in-the-app-store/",
month = sep,
year = {2024},
note = "Accessed: 2025-4-3"
}

@misc{web-appStoreCount,
author = {Statista},
title = {Biggest app stores in the world 2024},
url = "https://www.statista.com/statistics/276623/number-of-apps-available-in-leading-app-stores/",
month = aug,
year = {2024},
note = "Accessed 2025-4-3"
}

@article{AbuFarha2024,
  title = {Drivers and outcomes of a shopper-retailer’s app relationship},
  volume = {81},
  ISSN = {0969-6989},
  url = {http://dx.doi.org/10.1016/j.jretconser.2024.104002},
  DOI = {10.1016/j.jretconser.2024.104002},
  journal = {Journal of Retailing and Consumer Services},
  publisher = {Elsevier BV},
  author = {Abu Farha,  Allam K. and El Hedhli,  Kamel and Alnawas,  Ibrahim and Zourrig,  Haithem and Becheur,  Imene},
  year = {2024},
  month = nov,
  pages = {104002}
}

@inproceedings{Xanthopoulos2013,
author = {Xanthopoulos, Spyros and Xinogalos, Stelios},
title = {A comparative analysis of cross-platform development approaches for mobile applications},
year = {2013},
isbn = {9781450318518},
publisher = {Association for Computing Machinery},
address = {New York, NY, USA},
url = {https://doi.org/10.1145/2490257.2490292},
doi = {10.1145/2490257.2490292},
booktitle = {Proceedings of the 6th Balkan Conference in Informatics},
pages = {213–220},
numpages = {8},
location = {Thessaloniki, Greece},
series = {BCI '13}
}

@inproceedings{Oltrogge2018,
  title = {The Rise of the Citizen Developer: Assessing the Security Impact of Online App Generators},
  url = {http://dx.doi.org/10.1109/SP.2018.00005},
  DOI = {10.1109/sp.2018.00005},
  booktitle = {2018 IEEE Symposium on Security and Privacy (SP)},
  publisher = {IEEE},
  author = {Oltrogge,  Marten and Derr,  Erik and Stransky,  Christian and Acar,  Yasemin and Fahl,  Sascha and Rossow,  Christian and Pellegrino,  Giancarlo and Bugiel,  Sven and Backes,  Michael},
  year = {2018},
  month = may 
}

@book{walter2015learning,
  title={Learning MIT App Inventor: A Hands-On Guide to Building Your Own Android Apps},
  author={Walter, D. and Sherman, M.},
  isbn={9780133798630},
  lccn={2014950962},
  series={Addison-Wesley learning series},
  year={2015},
  publisher={Addison-Wesley}
}

@inproceedings{Azmy2023,
author = {Azmy, Imran Harith and Azmi, Azri and Kama, Nazri and Mohd Rusli, Hazlifah and Chuprat, Suriayati and Anuar, Asyraf Wahi},
title = {Methods for Application Development by Non-Programmers: A Systematic Literature Review},
year = {2023},
isbn = {9798400708053},
publisher = {Association for Computing Machinery},
address = {New York, NY, USA},
url = {https://doi.org/10.1145/3631991.3631992},
doi = {10.1145/3631991.3631992},
booktitle = {Proceedings of the 2023 5th World Symposium on Software Engineering},
pages = {1–8},
numpages = {8},
location = {Tokyo, Japan},
series = {WSSE '23}
}

@inproceedings{Nunkesser2018,
author = {Nunkesser, Robin},
title = {Beyond web/native/hybrid: a new taxonomy for mobile app development},
year = {2018},
isbn = {9781450357128},
publisher = {Association for Computing Machinery},
address = {New York, NY, USA},
url = {https://doi.org/10.1145/3197231.3197260},
doi = {10.1145/3197231.3197260},
booktitle = {Proceedings of the 5th International Conference on Mobile Software Engineering and Systems},
pages = {214–218},
numpages = {5},
location = {Gothenburg, Sweden},
series = {MOBILESoft '18}
}

@inbook{Vinson2008,
  title = {A Practical Guide to Ethical Research Involving Humans},
  ISBN = {9781848000445},
  url = {http://dx.doi.org/10.1007/978-1-84800-044-5_9},
  DOI = {10.1007/978-1-84800-044-5_9},
  booktitle = {Guide to Advanced Empirical Software Engineering},
  publisher = {Springer London},
  author = {Vinson,  Norman G. and Singer,  Janice},
  year = {2008},
  pages = {229–256}
}

@article{Braun2006,
  title = {Using thematic analysis in psychology},
  volume = {3},
  ISSN = {1478-0895},
  url = {http://dx.doi.org/10.1191/1478088706qp063oa},
  DOI = {10.1191/1478088706qp063oa},
  number = {2},
  journal = {Qualitative Research in Psychology},
  publisher = {Informa UK Limited},
  author = {Braun,  Virginia and Clarke,  Victoria},
  year = {2006},
  month = jan,
  pages = {77–101}
}

@article{Kitchenham2002,
author = {Kitchenham, Barbara A. and Pfleeger, Shari Lawrence},
title = {Principles of survey research: part 3: constructing a survey instrument},
year = {2002},
issue_date = {March 2002},
publisher = {Association for Computing Machinery},
address = {New York, NY, USA},
volume = {27},
number = {2},
issn = {0163-5948},
url = {https://doi.org/10.1145/511152.511155},
doi = {10.1145/511152.511155},
journal = {SIGSOFT Softw. Eng. Notes},
month = mar,
pages = {20–24},
numpages = {5},
}

@inproceedings{Peruma2024,
  title = {A Developer-Centric Study Exploring Mobile Application Security Practices and Challenges},
  url = {http://dx.doi.org/10.1109/ICSME58944.2024.00081},
  DOI = {10.1109/icsme58944.2024.00081},
  booktitle = {2024 IEEE International Conference on Software Maintenance and Evolution (ICSME)},
  publisher = {IEEE},
  author = {Peruma,  Anthony and Huo,  Timothy and Araújo,  Ana Catarina and Imanmka,  Jake and Kazman,  Rick},
  year = {2024},
  month = oct,
  pages = {778–790}
}

@article{Tong2007,
  title = {Consolidated criteria for reporting qualitative research (COREQ): a 32-item checklist for interviews and focus groups},
  volume = {19},
  ISSN = {1464-3677},
  url = {http://dx.doi.org/10.1093/intqhc/mzm042},
  DOI = {10.1093/intqhc/mzm042},
  number = {6},
  journal = {International Journal for Quality in Health Care},
  publisher = {Oxford University Press (OUP)},
  author = {Tong,  A. and Sainsbury,  P. and Craig,  J.},
  year = {2007},
  month = sep,
  pages = {349–357}
}

@article{Lenberg2023,
  title = {Qualitative software engineering research: Reflections and guidelines},
  volume = {36},
  ISSN = {2047-7481},
  url = {http://dx.doi.org/10.1002/smr.2607},
  DOI = {10.1002/smr.2607},
  number = {6},
  journal = {Journal of Software: Evolution and Process},
  publisher = {Wiley},
  author = {Lenberg,  Per and Feldt,  Robert and Gren,  Lucas and Wallgren Tengberg,  Lars G\"{o}ran and Tidefors,  Inga and Graziotin,  Daniel},
  year = {2023},
  month = sep 
}

@article{Lenberg2015,
  title = {Behavioral software engineering: A definition and systematic literature review},
  volume = {107},
  ISSN = {0164-1212},
  url = {http://dx.doi.org/10.1016/j.jss.2015.04.084},
  DOI = {10.1016/j.jss.2015.04.084},
  journal = {Journal of Systems and Software},
  publisher = {Elsevier BV},
  author = {Lenberg,  Per and Feldt,  Robert and Wallgren,  Lars G\"{o}ran},
  year = {2015},
  month = sep,
  pages = {15–37}
}

@article{Stol2018,
author = {Stol, Klaas-Jan and Fitzgerald, Brian},
title = {The ABC of Software Engineering Research},
year = {2018},
issue_date = {July 2018},
publisher = {Association for Computing Machinery},
address = {New York, NY, USA},
volume = {27},
number = {3},
issn = {1049-331X},
url = {https://doi.org/10.1145/3241743},
doi = {10.1145/3241743},
journal = {ACM Trans. Softw. Eng. Methodol.},
month = sep,
articleno = {11},
numpages = {51},
}

@article{Storey2020,
  title = {The who,  what,  how of software engineering research: a socio-technical framework},
  volume = {25},
  ISSN = {1573-7616},
  url = {http://dx.doi.org/10.1007/s10664-020-09858-z},
  DOI = {10.1007/s10664-020-09858-z},
  number = {5},
  journal = {Empirical Software Engineering},
  publisher = {Springer Science and Business Media LLC},
  author = {Storey,  Margaret-Anne and Ernst,  Neil A. and Williams,  Courtney and Kalliamvakou,  Eirini},
  year = {2020},
  month = aug,
  pages = {4097–4129}
}

@article{linaaker2015guidelines,
  title={Guidelines for conducting surveys in software engineering},
  author={Lin{\aa}ker, Johan and Sulaman, Sardar Muhammad and de Mello, Rafael Maiani and H{\"o}st, Martin},
  year={2015},
  publisher={Department of Computer Science, Lund University}
}

@misc{WHO_disability_report,
author = {World Health Organization},
title = {Disability},
howpublished = {\url{https://www.who.int/en/news-room/fact-sheets/detail/disability-and-health}},
month = mar,
year = {2023},
note = "Accessed 2025-04-03"
}

@article{hersh2019learning,
  title={Learning technology and disability—Overcoming barriers to inclusion: Evidence from a multicountry study},
  author={Hersh, Marion and Mouroutsou, Stella},
  journal={British Journal of Educational Technology},
  volume={50},
  number={6},
  pages={3329--3344},
  year={2019},
  publisher={Wiley Online Library}
}

@article{ross2020epidemiology,
  title={An epidemiology-inspired large-scale analysis of android app accessibility},
  author={Ross, Anne Spencer and Zhang, Xiaoyi and Fogarty, James and Wobbrock, Jacob O},
  journal={ACM Transactions on Accessible Computing (TACCESS)},
  year={2020},
  publisher={ACM New York, NY, USA},
  url={https://doi.org/10.1145/3348797}
}

@article{di2022making,
  title={The making of accessible android applications: an empirical study on the state of the practice},
  author={Di Gregorio, Marianna and Di Nucci, Dario and Palomba, Fabio and Vitiello, Giuliana},
  journal={Empirical Software Engineering},
  volume={27},
  number={6},
  pages={145},
  year={2022},
  publisher={Springer},
  url={https://doi.org/10.1007/s10664-022-10182-x}
}

@article{Paiva2021,
  title = {Accessibility and Software Engineering Processes: A Systematic Literature Review},
  volume = {171},
  ISSN = {0164-1212},
  url = {http://dx.doi.org/10.1016/j.jss.2020.110819},
  DOI = {10.1016/j.jss.2020.110819},
  journal = {Journal of Systems and Software},
  publisher = {Elsevier BV},
  author = {Paiva,  Débora Maria Barroso and Freire,  André Pimenta and de Mattos Fortes,  Renata Pontin},
  year = {2021},
  month = jan,
  pages = {110819}
}

@inproceedings{alshayban2020accessibility,
  title={Accessibility issues in Android apps: state of affairs, sentiments, and ways forward},
  author={Alshayban, Abdulaziz and Ahmed, Iftekhar and Malek, Sam},
  booktitle={Proceedings of the ACM/IEEE 42nd International Conference on Software Engineering},
  year={2020},
}

@inproceedings{vendome2019can,
  title={Can everyone use my app? an empirical study on accessibility in android apps},
  author={Vendome, Christopher and Solano, Diana and Li{\~n}{\'a}n, Santiago and Linares-V{\'a}squez, Mario},
  booktitle={2019 IEEE International Conference on Software Maintenance and Evolution},
  year={2019},
  organization={IEEE},
  doi={10.1109/ICSME.2019.00014},
}

@article{indika2024exploring,
  series = {HICSS},
  title = {Exploring Accessibility Trends and Challenges in Mobile App Development: A Study of Stack Overflow Questions},
  ISSN = {2572-6862},
  url = {http://dx.doi.org/10.24251/HICSS.2025.885},
  DOI = {10.24251/hicss.2025.885},
  booktitle = {Proceedings of the 57th Hawaii International Conference on System Sciences},
  publisher = {Hawaii International Conference on System Sciences},
  author = {Indika,  Amila and Lee,  Christopher and Wang,  Haochen and Lisoway,  Justin and Peruma,  Anthony and Kazman,  Rick},
  year = {2025},
  collection = {HICSS}
}

@article{ma2022first,
  title={A First Look at Dark Mode in Real-World Android Apps},
  author={Ma, Suyu and Chen, Chunyang and Khalajzadeh, Hourieh and Grundy, John},
  journal={ACM Transactions on Software Engineering and Methodology},
  year={2022},
  publisher={ACM New York, NY},
}

@misc{qualtrics,
    title={Qualtrics XM: The Leading Experience Management Software},
    author={},
    year={},
    howpublished={\url{https: //www.qualtrics.com/}}
}

@article{kitchenham2002principles,
  title={Principles of survey research: part 3: constructing a survey instrument},
  author={Kitchenham, Barbara A and Pfleeger, Shari Lawrence},
  journal={ACM SIGSOFT Software Engineering Notes},
  volume={27},
  number={2},
  pages={20--24},
  year={2002},
  publisher={ACM New York, NY, USA}
}

@misc{kasunic2005designing,
  title={Designing an effective survey},
  author={Kasunic, Mark},
  year={2005},
  publisher={Citeseer}
}

@misc{linkedin_stats,
    title={LinkedIn - statistics \& facts},
    author={Dixon, Stacy Jo},
    year={2024},
    howpublished={\url{https://www.statista.com/topics/951/linkedin/\#topicOverview}}
}

@article{mirabeau2013utility,
  title={The utility of using social media networks for data collection in survey research},
  author={Mirabeau, Laurent and Mignerat, Muriel and Grang{\'e}, Camille},
  year={2013}
}

@inproceedings{kanij2013lessons,
  title={Lessons learned from conducting industry surveys in software testing},
  author={Kanij, Tanjila and Merkel, Robert and Grundy, John},
  booktitle={2013 1st International Workshop on Conducting Empirical Studies in Industry (CESI)},
  pages={63--66},
  year={2013},
  organization={IEEE}
}

@article{wagner2020challenges,
  title={Challenges in survey research},
  author={Wagner, Stefan and Mendez, Daniel and Felderer, Michael and Graziotin, Daniel and Kalinowski, Marcos},
  journal={Contemporary Empirical Methods in Software Engineering},
  pages={93--125},
  year={2020},
  publisher={Springer}
}

@article{leite2021accessibility,
  title={Accessibility in the mobile development industry in Brazil: Awareness, knowledge, adoption, motivations and barriers},
  author={Leite, Manoel Victor Rodrigues and Scatalon, Lilian Passos and Freire, Andr{\'e} Pimenta and Eler, Marcelo Medeiros},
  journal={Journal of Systems and Software},
  volume={177},
  pages={110942},
  year={2021},
  publisher={Elsevier}
}

@article{chen2021accessible,
  title={Accessible or not? an empirical investigation of android app accessibility},
  author={Chen, Sen and Chen, Chunyang and Fan, Lingling and Fan, Mingming and Zhan, Xian and Liu, Yang},
  journal={IEEE Transactions on Software Engineering},
  volume={48},
  number={10},
  pages={3954--3968},
  year={2021},
  publisher={IEEE}
}

@inproceedings{Ballantyne2018,
  series = {MUM 2018},
  title = {Study of Accessibility Guidelines of Mobile Applications},
  url = {http://dx.doi.org/10.1145/3282894.3282921},
  DOI = {10.1145/3282894.3282921},
  booktitle = {Proceedings of the 17th International Conference on Mobile and Ubiquitous Multimedia},
  publisher = {ACM},
  author = {Ballantyne,  Mars and Jha,  Archit and Jacobsen,  Anna and Hawker,  J. Scott and El-Glaly,  Yasmine N.},
  year = {2018},
  month = nov,
  collection = {MUM 2018}
}

@article{yan2019current,
  title={The current status of accessibility in mobile apps},
  author={Yan, Shunguo and Ramachandran, PG},
  journal={ACM Transactions on Accessible Computing (TACCESS)},
  volume={12},
  number={1},
  pages={1--31},
  year={2019},
  publisher={ACM New York, NY, USA}
}

@article{oliveira2024exploring_1,
  title={Exploring the Influence of Software Evolution on Mobile App Accessibility: Insights from User Reviews},
  author={Oliveira, Alberto Dumont Alves and dos Santos, Paulo Sergio Henrique and Aljedaani, Wajdi and Eler, Marcelo Medeiros},
  journal={Journal of the Brazilian Computer Society},
  volume={30},
  number={1},
  pages={584--607},
  year={2024}
}

@article{acosta2021accessibility,
  title={Accessibility in native mobile applications for users with disabilities: A scoping review},
  author={Acosta-Vargas, Patricia and Salvador-Acosta, Bel{\'e}n and Salvador-Ullauri, Luis and Villegas-Ch, William and Gonzalez, Mario},
  journal={Applied Sciences},
  volume={11},
  number={12},
  pages={5707},
  year={2021},
  publisher={MDPI}
}

@inproceedings{oliveira2024exploring_2,
  title={Exploring Accessibility of Mobile Applications Through User Feedback: Insights from App Reviews in a Systematic Literature Review},
  author={Oliveira, Alberto Dumont Alves and Eler, Marcelo Medeiros},
  booktitle={Proceedings of the XXIII Brazilian Symposium on Human Factors in Computing Systems},
  pages={1--15},
  year={2024}
}

@article{Fawad2024,
  title = {Android Source Code Smells: A Systematic Literature Review},
  volume = {55},
  ISSN = {1097-024X},
  url = {http://dx.doi.org/10.1002/spe.3401},
  DOI = {10.1002/spe.3401},
  number = {5},
  journal = {Software: Practice and Experience},
  publisher = {Wiley},
  author = {Fawad,  Muhammad and Rasool,  Ghulam and Palma,  Francis},
  year = {2024},
  month = dec,
  pages = {847–882}
}

@article{Zhan2022,
  title = {Research on Third-Party Libraries in Android Apps: A Taxonomy and Systematic Literature Review},
  volume = {48},
  ISSN = {2326-3881},
  url = {http://dx.doi.org/10.1109/TSE.2021.3114381},
  DOI = {10.1109/tse.2021.3114381},
  number = {10},
  journal = {IEEE Transactions on Software Engineering},
  publisher = {Institute of Electrical and Electronics Engineers (IEEE)},
  author = {Zhan,  Xian and Liu,  Tianming and Fan,  Lingling and Li,  Li and Chen,  Sen and Luo,  Xiapu and Liu,  Yang},
  year = {2022},
  month = oct,
  pages = {4181–4213}
}

@article{Senanayake2023,
  title = {Android Source Code Vulnerability Detection: A Systematic Literature Review},
  volume = {55},
  ISSN = {1557-7341},
  url = {http://dx.doi.org/10.1145/3556974},
  DOI = {10.1145/3556974},
  number = {9},
  journal = {ACM Computing Surveys},
  publisher = {Association for Computing Machinery (ACM)},
  author = {Senanayake,  Janaka and Kalutarage,  Harsha and Al-Kadri,  Mhd Omar and Petrovski,  Andrei and Piras,  Luca},
  year = {2023},
  month = jan,
  pages = {1–37}
}

@article{Li2017,
  title = {Static analysis of android apps: A systematic literature review},
  volume = {88},
  ISSN = {0950-5849},
  url = {http://dx.doi.org/10.1016/j.infsof.2017.04.001},
  DOI = {10.1016/j.infsof.2017.04.001},
  journal = {Information and Software Technology},
  publisher = {Elsevier BV},
  author = {Li,  Li and Bissyandé,  Tegawendé F. and Papadakis,  Mike and Rasthofer,  Siegfried and Bartel,  Alexandre and Octeau,  Damien and Klein,  Jacques and Traon,  Le},
  year = {2017},
  month = aug,
  pages = {67–95}
}

@inproceedings{Joorabchi2013,
  title = {Real Challenges in Mobile App Development},
  url = {http://dx.doi.org/10.1109/ESEM.2013.9},
  DOI = {10.1109/esem.2013.9},
  booktitle = {2013 ACM / IEEE International Symposium on Empirical Software Engineering and Measurement},
  publisher = {IEEE},
  author = {Joorabchi,  Mona Erfani and Mesbah,  Ali and Kruchten,  Philippe},
  year = {2013},
  month = oct 
}

@inbook{Heitktter2013,
  title = {Evaluating Cross-Platform Development Approaches for Mobile Applications},
  ISBN = {9783642366086},
  ISSN = {1865-1356},
  url = {http://dx.doi.org/10.1007/978-3-642-36608-6_8},
  DOI = {10.1007/978-3-642-36608-6_8},
  booktitle = {Web Information Systems and Technologies},
  publisher = {Springer Berlin Heidelberg},
  author = {Heitk\"{o}tter,  Henning and Hanschke,  Sebastian and Majchrzak,  Tim A.},
  year = {2013},
  pages = {120–138}
}

@inproceedings{Ahmad2017,
  title = {Challenges of mobile applications development: Initial results},
  url = {http://dx.doi.org/10.1109/ICSESS.2017.8342956},
  DOI = {10.1109/icsess.2017.8342956},
  booktitle = {2017 8th IEEE International Conference on Software Engineering and Service Science (ICSESS)},
  publisher = {IEEE},
  author = {Ahmad,  Arshad and Feng,  Chong and Tao,  Ma and Yousif,  Abdallah and Ge,  Shi},
  year = {2017},
  month = nov,
  pages = {464–469}
}

@inproceedings{Ali2017,
  title = {Same App,  Different App Stores: A Comparative Study},
  url = {http://dx.doi.org/10.1109/MOBILESoft.2017.3},
  DOI = {10.1109/mobilesoft.2017.3},
  booktitle = {2017 IEEE/ACM 4th International Conference on Mobile Software Engineering and Systems (MOBILESoft)},
  publisher = {IEEE},
  author = {Ali,  Mohamed and Joorabchi,  Mona Erfani and Mesbah,  Ali},
  year = {2017},
  month = may,
  pages = {79–90}
}

@article{Hu2018,
  title = {Studying the consistency of star ratings and the complaints in 1 \& 2-star user reviews for top free cross-platform Android and iOS apps},
  volume = {23},
  ISSN = {1573-7616},
  url = {http://dx.doi.org/10.1007/s10664-018-9604-y},
  DOI = {10.1007/s10664-018-9604-y},
  number = {6},
  journal = {Empirical Software Engineering},
  publisher = {Springer Science and Business Media LLC},
  author = {Hu,  Hanyang and Bezemer,  Cor-Paul and Hassan,  Ahmed E.},
  year = {2018},
  month = mar,
  pages = {3442–3475}
}

@inproceedings{Domnguezlvarez2019,
  series = {ESEC/FSE ’19},
  title = {Release practices for iOS and Android apps},
  url = {http://dx.doi.org/10.1145/3340496.3342762},
  DOI = {10.1145/3340496.3342762},
  booktitle = {Proceedings of the 3rd ACM SIGSOFT International Workshop on App Market Analytics},
  publisher = {ACM},
  author = {Domínguez-Álvarez,  Daniel and Gorla,  Alessandra},
  year = {2019},
  month = aug,
  pages = {15–18},
  collection = {ESEC/FSE ’19}
}

@article{suvvari2023shift,
  title={Shift Left: Moving the Inclusion of Accessibility Functionalities to the Left in Agile Product Development Life Cycle},
  author={Suvvari, Sunil Kumar},
  journal={Journal of Computational Analysis and Applications},
  volume={31},
  number={4},
  year={2023}
}

@article{sklavenitis2025scoping,
  title={A Scoping Review and Assessment Framework for Technical Debt in the Development and Operation of AI/ML Competition Platforms},
  author={Sklavenitis, Dionysios and Kalles, Dimitris},
  journal={Applied Sciences},
  volume={15},
  number={13},
  pages={7165},
  year={2025},
  publisher={MDPI}
}

@article{churchill2018putting,
  title={Putting accessibility first},
  author={Churchill, Elizabeth F},
  journal={Interactions},
  volume={25},
  number={5},
  pages={24--25},
  year={2018},
  publisher={ACM New York, NY, USA}
}

@misc{ArtifactPackage,
author = {},
title = {Artifact Package},
howpublished = {\url{https://doi.org/10.5281/zenodo.18238885}},
month = {},
year = {2026},
}

@book{ernst2021,
  author    = {Ernst, Neil and Kazman, Rick and Delange, Julien},
  title     = {{Technical Debt in Practice: How to Find It and Fix It}},
  publisher = {MIT Press},
  year      = {2021},
  address   = {Cambridge, MA},
  isbn      = {978-0262542111},
}
\end{document}